\documentclass[usenatbib]{mn2e}
\usepackage{natbib}
\usepackage{mathtools}
\input psfig.sty
\usepackage{epsfig}
\usepackage{amsmath,amssymb}
\usepackage{psfig}
\usepackage{graphicx}
\usepackage{float}
\usepackage[a4paper,colorlinks=true,pdfstartview=FitV,linkcolor=red,citecolor=blue,urlcolor=magenta]{hyperref}

\newcommand{\kpc} {{\,\rm kpc}}

\newcommand{\kms}{{\,\rm {km\,s^{-1}} }}

\renewcommand{\vec}[1]{{\mathbf #1}}

\def \apjs  {ApJS}
\def \apjl  {ApJL}

\def \chisq  {\ifmmode  \chi^2   \else  $\chi^2$  \fi}  
\def \spose#1{\hbox  to 0pt{#1\hss}}  
\def \lta{\mathrel{\spose{\lower 3pt\hbox{$\sim$}}\raise  2.0pt\hbox{$<$}}}
\def \gta{\mathrel{\spose{\lower  3pt\hbox{$\sim$}}\raise 2.0pt\hbox{$>$}}}
\def \kms {\ifmmode  \,\rm km\,s^{-1} \else $\,\rm km\,s^{-1}  $ \fi }
\def \kpc {\ifmmode  {\rm kpc}  \else ${\rm  kpc}$ \fi  }  
\def \Msun {\ifmmode M_{\odot} \else $M_{\odot}$ \fi} 
\def \hMsun {\ifmmode h^{-1}\,\rm M_{\odot} \else $h^{-1}\,\rm M_{\odot}$ \fi}
\def \LCDM {\ifmmode \Lambda{\rm CDM} \else $\Lambda{\rm CDM}$ \fi}
\def \sig8 {\ifmmode \sigma_8 \else $\sigma_8$ \fi} 
\def \OmegaM {\ifmmode \Omega_{\rm M} \else $\Omega_{\rm M}$ \fi} 
\def \OmegaL {\ifmmode \Omega_{\rm \Lambda} \else $\Omega_{\rm \Lambda}$\fi} 
\def \Deltavir {\ifmmode \Delta_{\rm vir} \else $\Delta_{\rm vir}$ \fi}
\def \rhocrit {\ifmmode \rho_{\rm crit} \else $\rho_{\rm crit}$ \fi}
\def \rhou {\ifmmode \rho_{\rm u} \else $\rho_{\rm u}$ \fi}
\def \zc {\ifmmode z_{\rm c} \else $z_{\rm c}$ \fi}
\def \rhos {\ifmmode \rho_{\rm s} \else $\rho_{\rm s}$ \fi} 
\def \rs {\ifmmode r_{\rm s} \else $r_{\rm s}$ \fi} 
\def \cvir {\ifmmode c_{\rm vir} \else $c_{\rm vir}$ \fi} 
\def \Rvir {\ifmmode r_{\rm vir} \else $R_{\rm vir}$ \fi}
\def \Vvir {\ifmmode V_{\rm  vir} \else  $V_{\rm vir}$  \fi} 
\def \Mvir {\ifmmode M_{\rm  vir} \else $M_{\rm  vir}$ \fi}  
\def \Nvir {\ifmmode N_{\rm  vir} \else $N_{\rm  vir}$ \fi}  
\def \Jvir {\ifmmode J_{\rm vir} \else $J_{\rm vir}$ \fi} 
\def \Evir {\ifmmode E_{\rm vir} \else $E_{\rm vir}$ \fi} 
\def \lam {\ifmmode \lambda  \else $\lambda$ \fi} 
\def \lamp {\ifmmode \lambda^{\prime} \else $\lambda^{\prime}$  \fi} 
\def \Vmax {\ifmmode V_{\rm  max} \else  $V_{\rm max}$  \fi} 
\def \Mgas {\ifmmode M_{\rm gas} \else $M_{\rm gas}$ \fi} 
\def \Mcg {\ifmmode M_{\rm cg} \else $M_{\rm cg}$\fi} 
\def \Mhg {\ifmmode M_{\rm hg} \else $M_{\rm hg}$ \fi} 
\def \Mdisc {\ifmmode M_{\rm disc} \else $M_{\rm disc}$ \fi} 
\def \Md {\ifmmode M_{\rm d} \else $M_{\rm d}$ \fi} 
\def \Mdm {\ifmmode M_{\rm DM} \else $M_{\rm DM}$ \fi} 
\def \Mda {\ifmmode M_{\rm d,0\%} \else $M_{\rm d,0\%}$ \fi} 
\def \Mdb {\ifmmode M_{\rm d,20\%} \else $M_{\rm d,20\%}$ \fi} 
\def \Mdc {\ifmmode M_{\rm d,40\%} \else $M_{\rm d,40\%}$ \fi} 
\def \md {\ifmmode m_{\rm d} \else $m_{\rm d}$ \fi} 
\def \Mb {\ifmmode M_{\rm b} \else $M_{\rm b}$ \fi} 
\def \Mbh {\ifmmode M_{\rm b,pri} \else $M_{\rm b,pri}$ \fi} 
\def \Mbs {\ifmmode M_{\rm b,sat} \else $M_{\rm b,sat}$ \fi} 
\def \zo {\ifmmode z_{0} \else $z_{0}$ \fi} 
\def \rd {\ifmmode r_{\rm d} \else $r_{\rm d}$ \fi}
\def \rg {\ifmmode r_{\rm g} \else $r_{\rm g}$ \fi}
\def \rb {\ifmmode r_{\rm b} \else $r_{\rm b}$\fi}
\def \rs {\ifmmode r_{\rm s} \else $r_{\rm s}$\fi}
\def \rc {\ifmmode r_{\rm c} \else $r_{\rm c}$\fi}
\def \rvir {\ifmmode r_{\rm vir} \else $r_{\rm vir}$\fi}
\def \rbh {\ifmmode r_{\rm b,pri} \else $r_{\rm b,pri}$ \fi} 
\def \rbs {\ifmmode r_{\rm b,sat} \else $r_{\rm b,sat}$ \fi} 

\title[]{Effects of Coupled Dark Energy on the Milky Way and its Satellites}
\author[C. Penzo et al.]{Camilla Penzo$^{1,2}$\thanks{penzo@mpia.de}, Andrea V. Macci\`o$^1$, Marco Baldi$^{3,4,5}$, Luciano Casarini$^6$,
\newauthor{Jose O\~{n}orbe$^{1}$}\\
\\$^1$ Max-Planck-Institut f\"ur Astronomie, K\"onigstuhl 17, 69117 Heidelberg, Germany;
\\$^2$ Fellow of the International Max Planck Research School for Astronomy and Cosmic Physics at the University of Heidelberg;
\\$^3$ Dipartimento di Astronomia, Universit\`a di Bologna, Italy;
\\$^{4}$ INAF - Osservatorio Astronomico di Bologna, via Ranzani 1, I-40127 Bologna, Italy;
\\$^{5}$ INFN - Sezione di Bologna, viale Berti Pichat 6/2, I-40127 Bologna, Italy;
\\$^6$ Departamento de Fisica, Universidade Federal do Espirito Santo, Av. Fernando Ferrari 514, 29075-910 Vitoria (ES), Brazil
}

\begin{document}
\pagerange{\pageref{firstpage}--\pageref{lastpage}} \pubyear{---}
\maketitle
\label{firstpage}
\begin{abstract}

  We present the first numerical simulations
  in coupled dark energy cosmologies with  high enough resolution to investigate the
  effects of the coupling on galactic and sub-galactic scales.
  We  choose two  constant couplings  and a  time-varying
  coupling function and we run  simulations of three Milky-Way-size
  halos ($\sim$10$^{12}$M$_{\odot}$), a lower mass halo (6$\times$10$^{11}$M$_{\odot}$)  and  a dwarf
  galaxy halo (5$\times$10$^{9}$M$_{\odot}$). We resolve each halo with
  several millions dark matter particles.
  On all scales the coupling causes lower halo concentrations and
  a reduced number of substructures with respect to $\Lambda$CDM.
  We show that the reduced concentrations are not due to different formation
  times, but they are related to the extra terms that appear in the equations
  describing the gravitational dynamics.
  On the scale of the Milky Way satellites, we show that the lower concentrations
  can help in reconciling observed and simulated rotation curves, but
  the coupling values necessary to  have a  significant  difference
  from \LCDM are outside the current observational  constraints.
  On the other hand, if other modifications to the standard model
  allowing a higher coupling  (e.g. massive neutrinos) are considered,
  coupled dark energy can become an interesting scenario to alleviate
  the small-scale issues of the $\Lambda $CDM model.

\end{abstract}

\section{Introduction}
\label{intro}

\quad\quad Since the discovery of the accelerated expansion of the universe \citep{Rie98,Per99}, a Cosmological Constant $\Lambda$ has been the most widely accepted explanation for the required negative pressure. 
Together with cold dark matter, today the dark sector is accounting for about 95\% of the total energy density \citep{Pla15} and builds the foundations for the so-called $\Lambda$ Cold Dark Matter ($\Lambda$CDM) model. 
Despite the highly successful inflationary \LCDM paradigm, the fundamental problems associated with the introduction of a cosmological constant, namely \emph{fine-tuning} and \emph{coincidence} problems \citep{Wei89}, have served as motivations for alternative descriptions of the dark sector. 
Introducing a time evolving scalar field (dark energy) responsible for the negative pressure is the approach of quintessence models \citep{Wet88,Pee88} and has been one of the most popular generalizations for the cosmological constant in the last decade. 
Furthermore, given the currently still unknown nature of the dark sector, the possibility of a non-null coupling between dark matter and dark energy has been considered \citep{Wet95,And98,Ame00,Bil00,Zim01,Ame01,Far04,Gro04}. 
Given that in these models dark matter and dark energy density evolutions are strongly coupled, this would in turn alleviate the \emph{coincidence} problem \citep{Man03,Mat03}. 
The effects of such interaction might be seen on the Cosmic Microwave Background (CMB), on supernovae and on the growth of structures, as pointed out by \citet{Mat03,Ame04a,Ame04b,Koi05,Guo07} and many others. 
Structure formation has been as well investigated via numerical simulations by \citet{Mac04,Bal10,LiB11,Car14a} and their follow up works, where the statistical distribution of structures has been studied. 
Both \citet{Bal10} and \citet{Car14a} found that, when introducing a coupling between dark energy and dark matter, halo concentrations decrease. 
In this work, we run the first high resolution simulations on galactic scales in coupled dark energy cosmology. Our aim is to obtain high enough resolutions to investigate the properties of the dark matter distribution at sub-galactic scales, mass scales at which the effects of the coupling have not yet been studied. The subhalos that we are interested in will in turn be the hosts of dwarf galaxies and their properties can be compared with observations of satellite dwarf galaxies of both Milky Way and Andromeda. 
In fact, despite \LCDM predictions on large scales being in very good agreement with galaxy clustering surveys \citep{Jon09,Ala15}, on galactic scales challenges between \LCDM predictions and observations have appeared.

Firstly, the \emph{missing satellites} problem, i.e. overabundance of substructures in \LCDM Milky-Way size halo simulations when compared to observations of the Milky Way dwarf galaxies \citep{Kly99,Moo99}. On the other hand, as showed in \citet{Mad08} and \citet{Mac10}, accounting for the baryonic physics drastically reduces the number of visible satellites.
Secondly, the \emph{core/cusp} problem, namely the inconsistency between the constant density cores estimated from observations and the cuspy inner density profiles found in \LCDM simulations. See \citet{Flo94,Moo94,Die05,Gen09,Wal11,Agn12,Sal12}, but also \citet{Bos01,Swa03,Sim05}. While this inconsistency can be attributed to baryonic feedback processes \citep[e.g.][]{Gov12,DiC14,Ono15}, for the case of Milky Way satellites the baryonic explanation is not straightforward since these objects can be almost completely dark matter dominated.
\citet{Bal10} and \citet{Car14a} showed that for halos with $M\gtrsim10^{13}M_{\odot}$ the coupling between dark matter and dark energy produces density profiles that are less cuspy in the inner density regions, which can help alleviating the core/cusp problem. The aim of this work is to investigate whether this effect persists at much lower masses. Moreover, concentrations of the most massive subhalos orbiting around a \LCDM Milky-Way size halo seem to be too high to be hosting the brightest dwarf galaxies observed. This translates into a prediction from \LCDM numerical simulations for the existence of massive dark matter subhalos that seem to have failed at forming stars, and is known as the \emph{too big to fail} problem \citep{Boy11,Lov12,Ras12,Tol12}. 
Whether these issues bring serious challenges for the \LCDM model or whether they can entirely be treated by invoking baryonic physics is currently under debate. With this work we aim at studing the properties of halos and their sub-structures to determine whether coupled dark energy cosmologies can alleviate the aforementioned issues. In Section~\ref{models} we summarize the theoretical model behind coupled dark energy and we specify our choices of coupling functions. In Section~\ref{setup} we described the numerical methods used to produce initial conditions and the N-body codes to run the simulations. In Section~\ref{results} we show our simulations results, for both halos and subhalos. Finally, in Section~\ref{conclusions} we present our conclusions. 
\section{Cosmological Models}
\label{models}
\quad\quad We present a study focused on understanding the non-linear effects of coupled dark energy models on galactic scales. The models that we consider allow for an interaction between dark matter and dark energy \citep{Ame00,Bil00,Zim01,Gro04,Mac04,Bal10} and obey the following Lagrangian:
\begin{align}
\mathcal{L} = \frac{1}{16\pi G}R -\frac{1}{2}\partial^{\mu}\partial_{\mu}\phi - V(\phi) - m(\phi)\bar\psi\psi + \mathcal{L}_{kin}[\psi],
\end{align}
where the mass $m(\phi)$ of the dark matter field $\psi$ is a function of the dark energy scalar field $\phi$, and the $ \mathcal{L}_{kin}[\psi] $ term includes the kinetic part of the dark matter Lagrangian. The choice of $m(\phi)$ specifies the coupling and in our work we use:
\begin{align}
m(\phi) = m_0\mathrm{e}^{-\beta(\phi)\frac{\phi}{M_{\rm Pl}}},
\end{align}
where $m_0$ is the mass at $z=0$, $M_{\rm Pl} \equiv 1/\sqrt{8\pi G}$ is the Planck mass, with $G$ being the Newton's constant, and $\beta(\phi)$ is the coupling function. The respective continuity equations for cold dark matter and dark energy are:
\begin{align}
\dot\rho_c + 3 H \rho_c &= -\beta(\phi)\dot\phi\rho_c,\\
\dot\rho_{\phi} + 3 H \rho_{\phi} &= +\beta(\phi)\dot\phi\rho_c, \nonumber,
\end{align}
where $\rho_c$ is the cold dark matter density and $\rho_{\phi}$ is the dark energy density, which is $\rho_{\phi} \equiv \dot\phi^2 + V(\phi)$, and $H\equiv \dot a / a$ is the Hubble parameter. Our choice for the self-interacting dark energy potential is $V(\phi) \propto e^{-\alpha\phi}$, with $\alpha = 0.08$.
 
The evolution of cold dark matter density perturbations is regulated by the following equation:
\begin{align}
\label{linper}
\ddot \delta_c + (2H - \beta \dot\phi) \dot\delta_c - \frac{3}{2} H^2 \left[ (1 + 2\beta^2) \Omega_c \delta_c + \Omega_b \delta_b \right] = 0,
\end{align}
where $\Omega_c$ and $\Omega_b$ are respectively the density parameters $\Omega_i \equiv \rho_i /\rho_{\rm crit}$ for cold dark matter and baryons, and $\rho_{\rm crit} = 8\pi G / 3H^2$. Two extra terms appear in Eq.~\ref{linper} compared to the \LCDM case: a friction term $-\beta\dot\phi\dot\delta_c$ and the factor $(1 + 2\beta^2)$ responsible for the enhancement of the gravitational force acting on cold dark matter particles, which is known as ``fifth force''. As pointed out in \citet{Bal11b}, in the linear regime both these extra terms produce an acceleration of growth of cold dark matter density perturbations. On the other hand, when considering the non-linear effects, the friction term is responsible for lowering the concentration of dark matter halos.

The appearance of extra terms becomes clear when calculating the acceleration felt by the $i$-th dark matter particle in a coupled dark energy cosmology for the limit of a light scalar field (see \citealt{Bal10} for calculation):
\begin{align}
\dot{\bar {v_i}} = \beta(\phi)\dot\phi\bar v_i + G[1+2\beta(\phi)^2]\sum_{j\neq i}\frac{m_j\bar r_{ij}}{|\bar r_{ij}|^3} \,.
\end{align}
The term $\beta(\phi)\dot\phi\bar v_i$ accelerates dark matter particles in the direction of their motion and thus lowers halo concentrations.

Based on \citet{Bal10} and \citet{Bal11b}, we chose three coupling scenarios. EXP003 and EXP008e3 are observationally viable models (see \citealt{Pet12} for CMB constraints on the coupling value), respectively with a constant coupling $\beta = 0.15$ and a coupling $\beta(\phi)$ varying with redshift (for more details on these models we refer to \citealt{Bal11a}). We also explore an extreme constant coupling case EXP006 with $\beta = 0.3$ (about 6$\sigma$ outside present observational limits, \citealt{Pet12}) to better understand its implications. 
\section{Numerical methods}
\label{setup}
\subsection{Initial Conditions and Coupled Dark Energy}
\begin{figure}   
\psfig{file=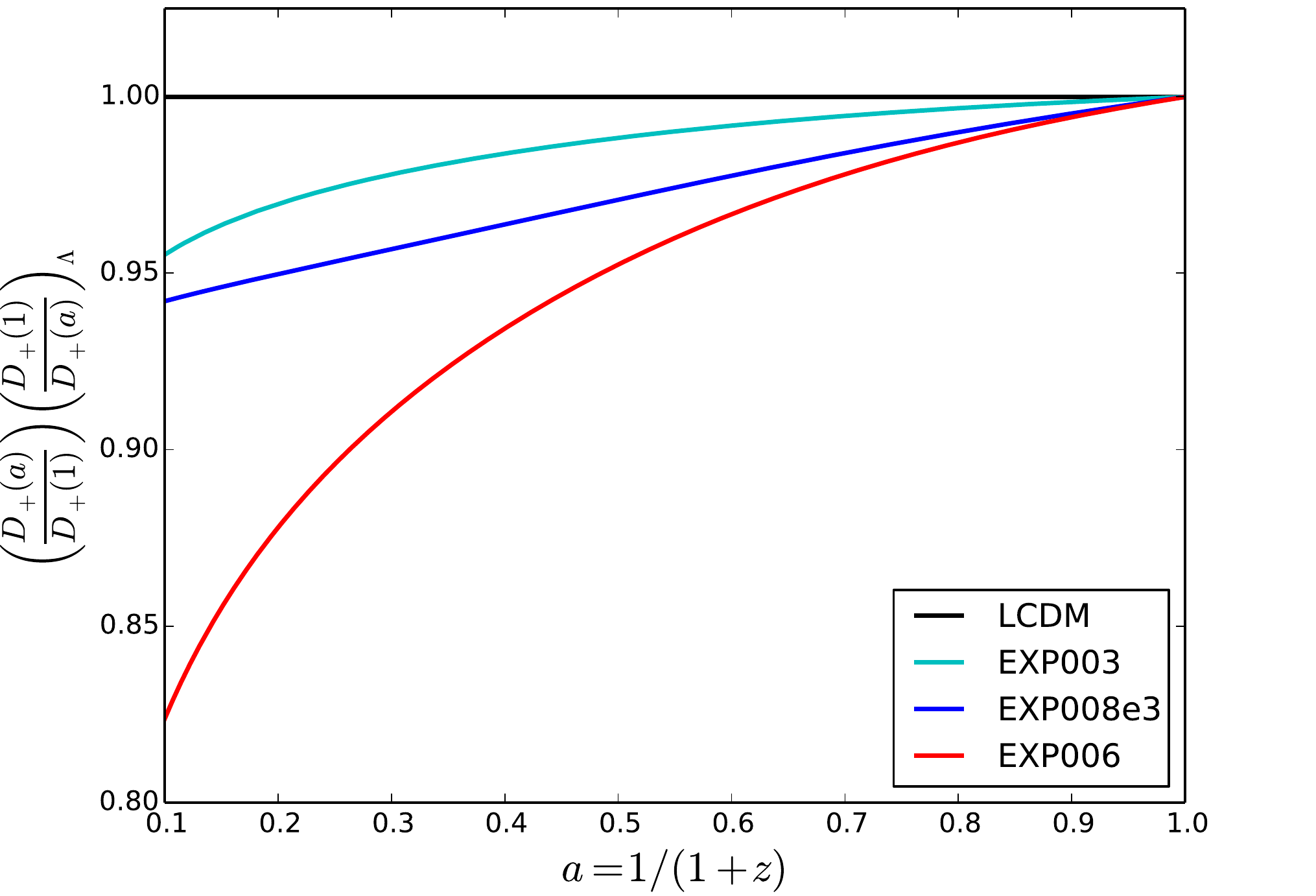,width=0.52\textwidth}
\caption{Linear growth factor evolutions for all cosmologies normalized to today's values divided by the \LCDM evolution.}
\label{fig:Dplus}
\end{figure}

\quad\quad As in \citet{Pen14}, we used {\sc grafic-de}, an extension of the initial condition generator {\sc   grafic-2} \citep{Ber01} such that initial conditions for a generic cosmological model can be produced once  the evolution of the  cosmological parameters are given as  an input. {\sc grafic-de} requires transfer functions, evolution of  the density parameters  $\Omega_i$, linear growth factor D$_{+}$ and f$_{\Omega}$, the logarithmic derivative of the growth factor with respect to the scale factor. As the original code, {\sc grafic-de} is  able to generate multi mass initial conditions from a cosmological box. In Fig.~\ref{fig:Dplus} we show the evolution of the linear growth factor D$_{+}$ for all four cosmological models. 
\begin{figure}   
\psfig{file=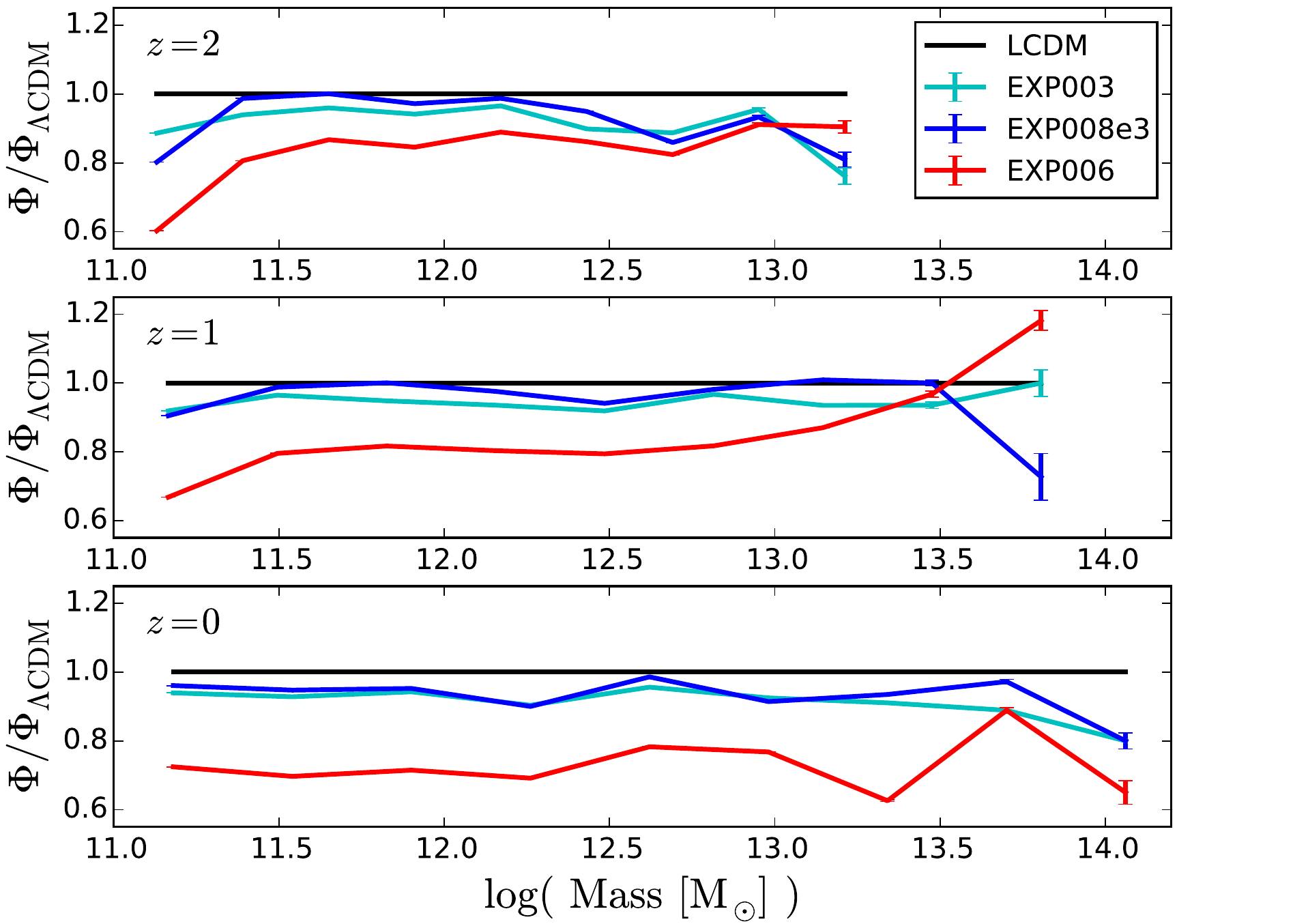,width=0.52\textwidth}
\caption{Ratio between the mass functions for the coupled dark energy cosmologies and the one from \LCDM for redshifts $z=2,1,0$ for the cosmological boxes of size 80 Mpc/h.}
\label{fig:ratiomf}
\end{figure}
The transfer functions for \LCDM have been produced using {\sc camb} \citep{Lew02}, while the transfer functions for the coupled dark energy models have been produced by weighting the \LCDM transfer functions with the D$_{+}$ of the coupled model. All initial conditions were created using the same random seeds, in order to be able to identify structures among the models. All the models share the same cosmological parameters and have at $z=0$: $\Omega_{b_0} = 0.0458$, $\Omega_{\rm DM_0} = 0.229$, $\rm H_0 = 70.2$ km s$^{-1}$ Mpc$^{-1}$, $\sigma_8$ = 0.816, $n_s = 0.968$, where these parameters are density parameters for baryons and dark matter, Hubble constant, root mean square of the fluctuation amplitudes and primeval spectra index.

\subsection{N-body Simulations}
\label{nbodysim}
\quad\quad We first generate two sets of uniform particle distributions, a 80 Mpc/h box and a 12 Mpc/h box, both with $350^3$ particles. The initial conditions were evolved with the code {\sc gadget-2} \citep{Spr05}, which includes the coupled dark energy implementation introduced in \citet{Bal10}. In Fig.~\ref{fig:ratiomf} we show the ratio between the mass functions for the coupled dark energy cosmologies with the one from \LCDM for redshifts $z=2,1,0$ for our 80 Mpc/h boxes.

We chose four dark matter halos in the \LCDM 80 Mpc/h box and one dwarf halo in the \LCDM 12 Mpc/h box, and looked for their corresponding realizations in the coupled dark energy simulations. Note that we used the same random seed for all initial conditions to be able to follow the same halos in all cosmological boxes. Our halos have been chosen so that no other halos with comparable masses were found within four of their virial radii. We then re-ran the cosmological boxes with increased resolution in a Lagrangian volume that includes all particles that at $z=0$ were found in three virial radii of each selected halo. 

\begin{figure*}
\centering
\begin{minipage}{180mm}
\psfig{file=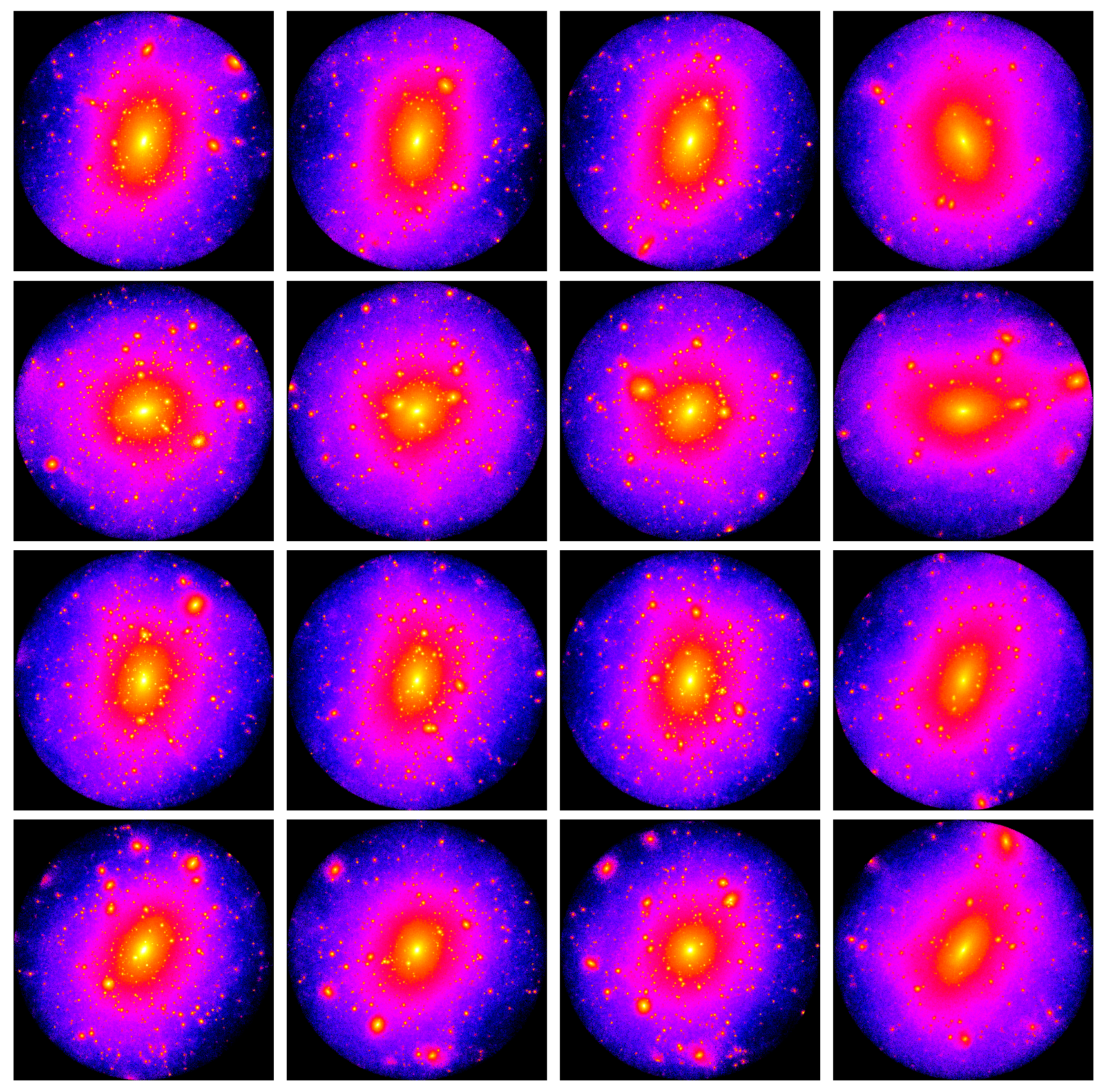,width=1.0\textwidth} 
\end{minipage}
\caption{Projected density maps of our sample at $z=0$. From first row to last we are showing halo$\alpha$, halo$\beta$, halo$\gamma$,halo$\delta$; from first column to last we are showing $\Lambda$CDM, EXP003, EXP008e3 and EXP006 cosmologies. All images are spheres of radius R$_{200}$, radius at which the density is equal to 200 times the critical density.}
\label{fig:maps}
\end{figure*}

Our final sample is composed of three Milky-Way-size halos (halo$\alpha$, halo$\beta$ and halo$\gamma$), a 6$\times$10$^{11}$M$_{\odot}$ halo (halo$\delta$) and a dwarf halo (halo$\epsilon$). For more details on the halos properties at $z=0$ see Table~\ref{tab:param}. For the halo identification we used the code Amiga Halo Finder \citep{Kno09}. The softening lengths are chosen to be $1/40$ of the intra-particle distance in the low resolution simulation divided by the refinement factor $\rm RF$; $\rm RF = 15$ for halo$\alpha$, halo$\beta$ and halo$\epsilon$, $\rm RF=24$ for halo$\gamma$ and halo$\delta$. Precisely, the softening lengths are 0.54 kpc for halo$\alpha$ and halo$\beta$, 0.34 kpc for halo$\gamma$ and halo$\delta$, 0.081 kpc for halo$\epsilon$. The particle masses at $z=0$ in the high resolution volumes are 3.8$\times 10^{5}$M$_{\odot}$ for halo$\alpha$ and halo$\beta$, 9.4$\times 10^{4}$M$_{\odot}$ for halo$\gamma$ and halo$\delta$, 1.3$\times 10^{3}$M$_{\odot}$ for halo$\epsilon$.
In Fig.~\ref{fig:maps} we show the projected density maps of the four most massive halos for each cosmological model. We will discuss the dwarf halo in Section~\ref{dwarfsection}. For the density maps and throughout the paper we chose to calculate halo properties using R$_{200}$, radius at which the density is 200 times the critical density $\rho_{c}$, with $\rho_{c}\equiv 3H^2 / (8\pi G)$.\smallbreak
\begin{table}
  \centering
  \caption{Physical properties of the five halos in all cosmologies, $\Lambda$CDM, EXP003, EXP008e3 and EXP006. We show mass at R$_{200}$, R$_{200}$, concentrations and number of particles within R$_{200}$.}
  \begin{tabular}{@{}lccccc@{}}
  \hline 
  &  M$_{\rm 200}$ &  R$_{\rm 200}$ &  c $\equiv$ R$_{\rm 200}$/r$_{\rm s}$ & N$_{\rm 200}$  \\ 
  & [\Msun] & [kpc] &  &\\
  \hline %
  \textbf{halo$\alpha$} \\ 
  \LCDM 		& 2.6$\times$10$^{12}$  &  284  &   11.8	&  6.8$\times$10$^{6}$     \\  
  EXP003 		& 2.5$\times$10$^{12}$  &  281  &   9.1	&  6.6$\times$10$^{6}$     \\  
  EXP008e3 	& 2.6$\times$10$^{12}$  &  282  &   10.2	&   6.7$\times$10$^{6}$    \\ 
  EXP006 		& 2.1$\times$10$^{12}$  &  265  &   4.6	&   5.5$\times$10$^{6}$    \\  
\hline   %
  \textbf{halo$\beta$} \\
  \LCDM 		& 2.5$\times$10$^{12}$  &   278  &   10.7	&   6.3$\times$10$^{6}$   \\  
  EXP003 		& 2.2$\times$10$^{12}$  &   267  &   8.0	&   5.7$\times$10$^{6}$    \\
  EXP008e3 	& 2.2$\times$10$^{12}$  &   268  &   8.7	&   5.8$\times$10$^{6}$    \\
  EXP006 		& 1.7$\times$10$^{12}$  &   246  &   4.3	&   4.5$\times$10$^{6}$    \\ 
  \hline
  \textbf{halo$\gamma$} \\
  \LCDM 		& 9.7$\times$10$^{11}$  &   204  &   10.8	&  1.0$\times$10$^{7}$     \\ 
  EXP003 		& 9.3$\times$10$^{11}$  &   201  &   8.6	&   9.9$\times$10$^{6}$    \\ 
  EXP008e3 	& 9.5$\times$10$^{11}$  &   203  &   9.6	&   1.0$\times$10$^{7}$    \\
  EXP006 		& 7.6$\times$10$^{11}$  &   188  &   3.2	&    8.1$\times$10$^{6}$   \\
  \hline
  \textbf{halo$\delta$} \\
  \LCDM 		& 6.4$\times$10$^{11}$  &  177  &   13.3	&    6.8$\times$10$^{6}$    \\
  EXP003 		& 5.6$\times$10$^{11}$  &  170  &   9.7	&    6.0$\times$10$^{6}$    \\
  EXP008e3 	& 5.9$\times$10$^{11}$  &  172  &   11.0	&    6.3$\times$10$^{6}$   \\
  EXP006 		& 5.3$\times$10$^{11}$  &  166  &   4.7	&    5.6$\times$10$^{6}$     \\ 
  \hline
    \textbf{halo$\epsilon$} \\
  \LCDM 		& 4$\times$10$^{9}$   & 33   & 15.3	&   3.1$\times$10$^{6}$	           \\  
  EXP006 		& 3$\times$10$^{9}$   & 30   &  6.6	&   2.4$\times$10$^{6}$	         \\  
  \hline
  \end{tabular}
  \label{tab:param}
\end{table} 

\section{Results}
\label{results}
\subsection{Host Halos Properties}
We first analyze the properties of the four most massive halos, halo$\alpha$, halo$\beta$, halo$\gamma$ and halo$\delta$. We show concentrations, density profiles, rotation curves, evolution of the scale radius and accretion histories.
\subsubsection{Concentrations and Density Profiles}
\label{concsec}
\quad\quad By introducing a coupling between dark matter and dark energy, halo concentrations decrease. This was also shown in \citealt{Bal10}, \citealt{LiB11}, and \citealt{Car14a} for halos with masses $M\gtrsim10^{13}M_{\odot}$. In this work we investigate mass scales  $M\lesssim10^{12}M_{\odot}$. The resolution that we are able to reach is higher thanks to the multi-mass technique. In Table~\ref{tab:param} we show the concentration values for each halo, for which we use the definition 
\begin{align}
c \equiv R_{200} / r_s,
\end{align}
where R$_{200}$ is the radius at which the density equals 200 times the critical density and r$_s$ is the scale radius in the Navarro Frenk and White (NFW) profile \citep{Nav97}. We computed r$_s$  via a $\chi^2$ minimization procedure using the Levenberg \& Marquart method as in \citet{Mac08}. In agreement with literature, we find that halos which lived in a coupled dark energy cosmology have lower concentrations.
\begin{figure*}
\begin{minipage}{180mm}
\centering   
\psfig{file=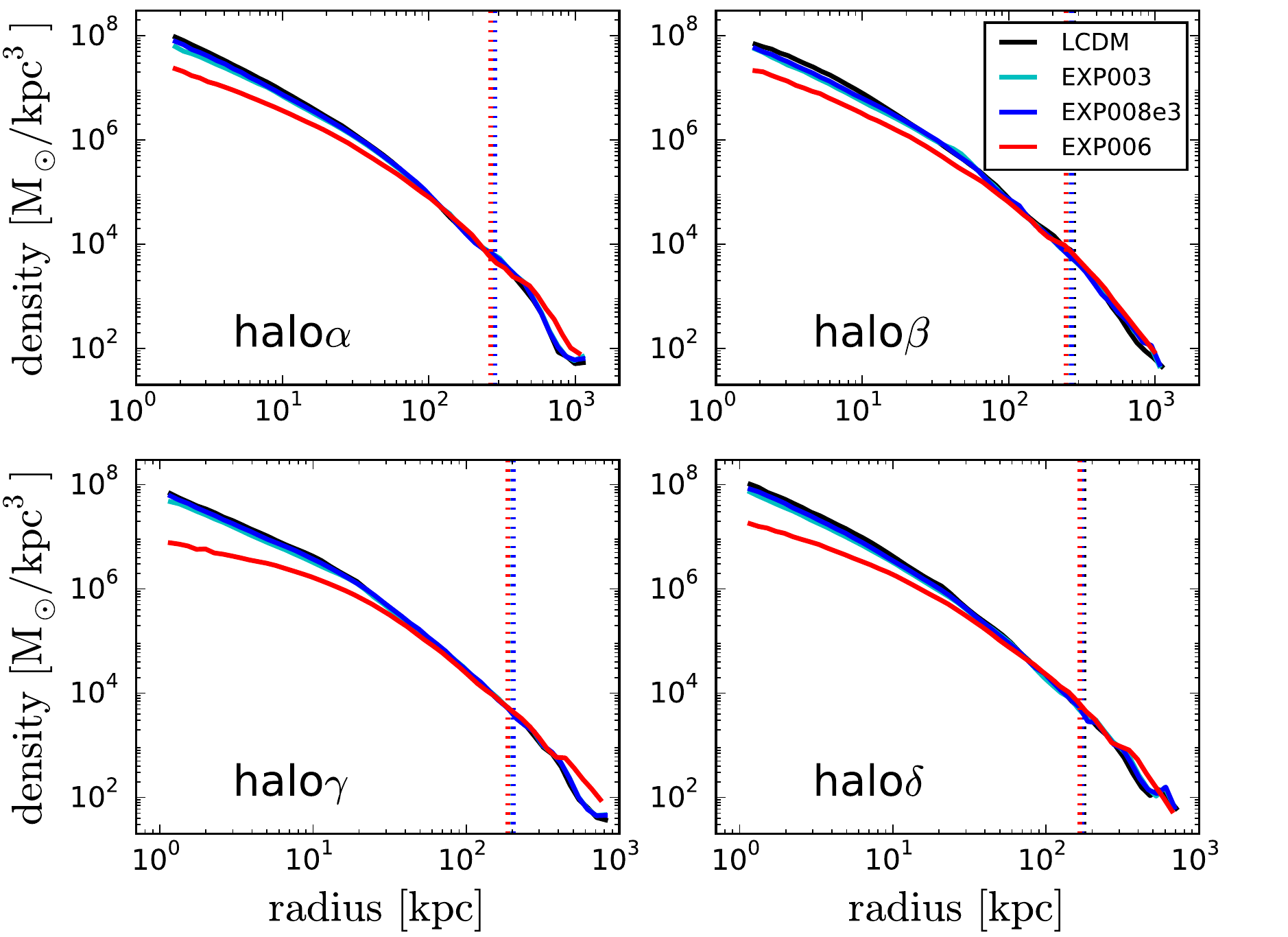,width=0.8\textwidth}
\caption{Density profiles for halo$\alpha$, halo$\beta$, halo$\gamma$ and halo$\delta$ at $z=0$, each for \LCDM (black), EXP003 (cyan), EXP008e3 (blue) and EXP006 (red). The inner radius is equal to three times the softening length, while the outer radius is four times R$_{200}$ of each halo. The vertical dashed lines mark R$_{200}$ for each halo in each cosmology.}
\label{fig:prof}
\end{minipage}
\end{figure*}
Fig.~\ref{fig:prof} shows the density profiles for halo$\alpha$, halo$\beta$, halo$\gamma$ and halo$\delta$. The behavior of the profiles as a function of cosmology is maintained for all four halos, with a significant flattening of the inner part of the profiles only for the extreme coupled cosmology EXP006, while differences are less evident in the EXP003 and EXP008e3 halos. Interestingly, the EXP006 realization of halo$\gamma$ (M = 7.6$\times$10$^{11}$M$_{\odot}$) produces a much flatter halo profile, with slope $\alpha$=-0.8, which falls out of NFW parametrization. On the other hand, all other profiles of halo$\alpha$, halo$\beta$, halo$\gamma$ and halo$\delta$ in all cosmologies are well described by the NFW profile.

\begin{figure}   
\psfig{file=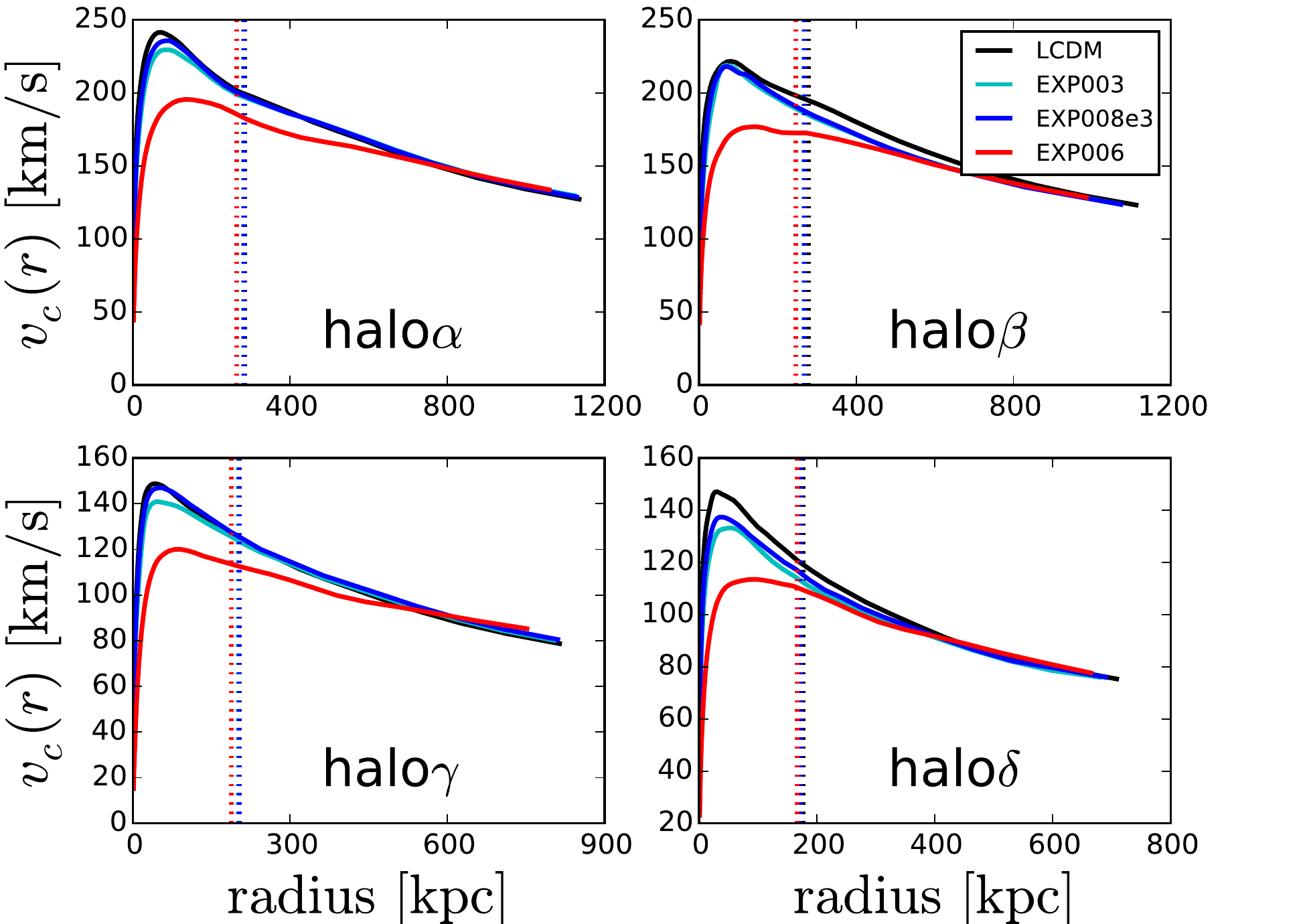,width=0.52\textwidth}
\caption{Rotation curves for halo$\alpha$, halo$\beta$, halo$\gamma$ and halo$\delta$ at $z=0$, each for \LCDM, EXP003, EXP008e3 and EXP006. The inner radius is equal to three times the softening length, while the outer radius is four times R$_{200}$ of each halo. The vertical dashed lines mark R$_{200}$ for each halo in each cosmology.}
\label{fig:curv}
\end{figure}

Additionally, in Fig.~\ref{fig:curv} we show the rotation curves at $z=0$ for the four halos. For models within the observational constraints the rotation curves are not significantly affected. The only case in which we observe a considerable flattening is the extreme model EXP006, for all four cases. This is in agreement with \citet{Pen14}, where we find that differences in rotation curves among models within observational constraints for dynamical dark energy are not significant in the dark matter only case. On the contrary, in hydrodynamical simulations we find observable differences in rotation curves due to the effects of baryons which enhance the variations in the dark matter accretion. We expect the same enhancement also in coupled dark energy models once hydrodynamics is taken into account. This aspect will be explored in a future work.
\subsubsection{NFW Scale Radius Evolution}
\begin{figure}   
\centering
\psfig{file=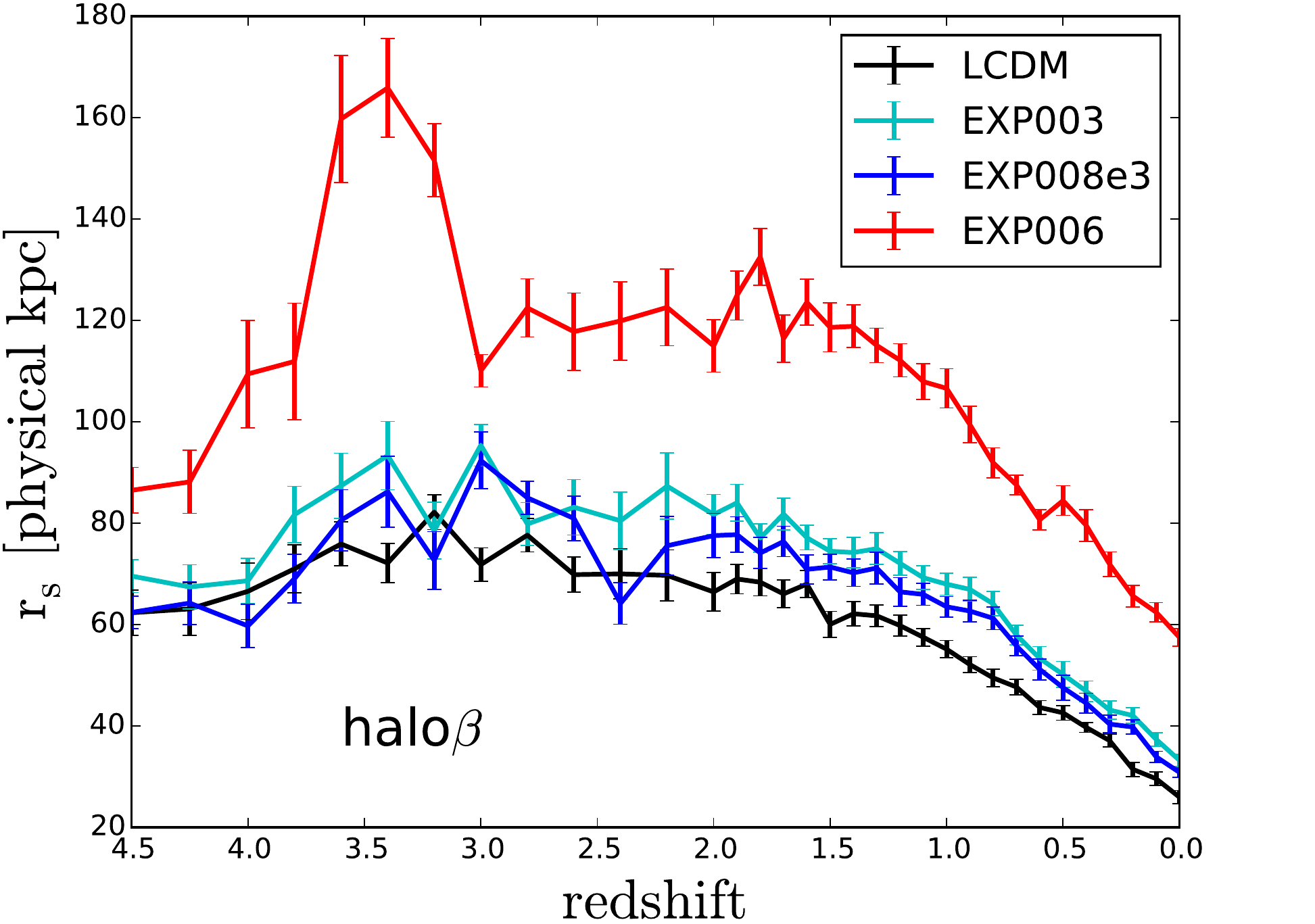,width=0.45\textwidth}
\caption{Scale radius obtained by fitting an NFW density profile using the Levenberg \& Marquart method for halo$\beta$ as a function of redshift.}
\label{fig:rs}
\end{figure}

\quad\quad In the Section~\ref{concsec} we have showed that halos that form in a coupled dark energy cosmology with a high value for the coupling constant have concentrations that are significantly lower at $z=0$. Given that almost all halos are well described by a NFW density profile, it means that their NFW scale radii $r_s$ are much larger than the scale radii of the corresponding \LCDM realizations. In Fig.~\ref{fig:rs} we show the behavior of the scale radius $r_s$ as a function of redshift for halo$\beta$ in all four cosmologies; the other Milky-Way size halos have similar behaviors. Compared to the \LCDM case, halos which live in coupled dark energy cosmologies show a larger scale radius at all redshifts.
\subsubsection{Main Halos Accretion Histories}
\label{acc_hist}
\begin{figure}   
\psfig{file=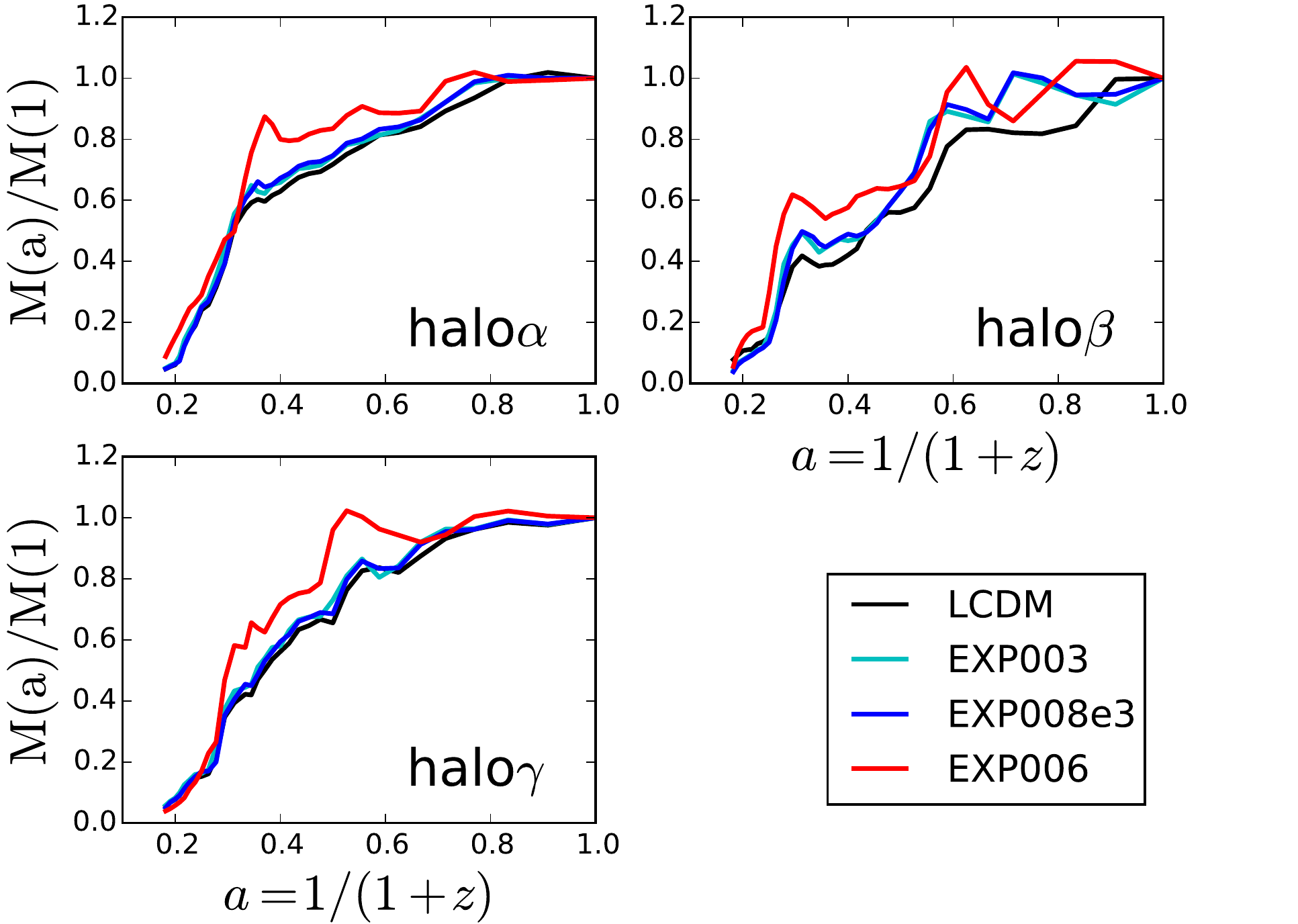,width=0.52\textwidth}
\caption{Evolution of the mass enclosed in R$_{200}$ normalized to the mass at $a=1$ as a function of the scale factor for halo$\alpha$, halo$\beta$, halo$\gamma$ in all four cosmologies.}
\label{fig:acc}
\end{figure}

\quad\quad In order to investigate the origin of the different concentrations, especially in the EXP006 cosmology, in this Section we concentrate on the halo formation times. Firstly, in Fig.~\ref{fig:acc} we show the accretion histories, namely the evolution of the mass enclosed in R$_{200}$ normalized to the value of the mass at $z=0$ as a function of expansion factor $a = 1/(1+z)$. 
Halos growing in $\Lambda{\rm CDM}$, EXP003 and EXP008e3 cosmologies show similar accretion histories especially for halo$\alpha$ and halo$\gamma$; while halos living in the EXP006 cosmology accrete their mass earlier on compared to their \LCDM realizations. Among the three halos, coupled cosmologies runs show unexpected drops in the accretion histories. These would be unusual in a \LCDM scenario since halo total masses do not decrease unless it is a temporary effect of a merger (see for instance halo$\alpha$ and halo$\beta$ around $a=0.3$). On the other hand, in the case of coupled cosmologies the friction term in Eq.~\ref{linper} is responsible for injecting kinetic energy into the system, which may cause some particles to become gravitationally unbound.

\begin{figure}   
\psfig{file=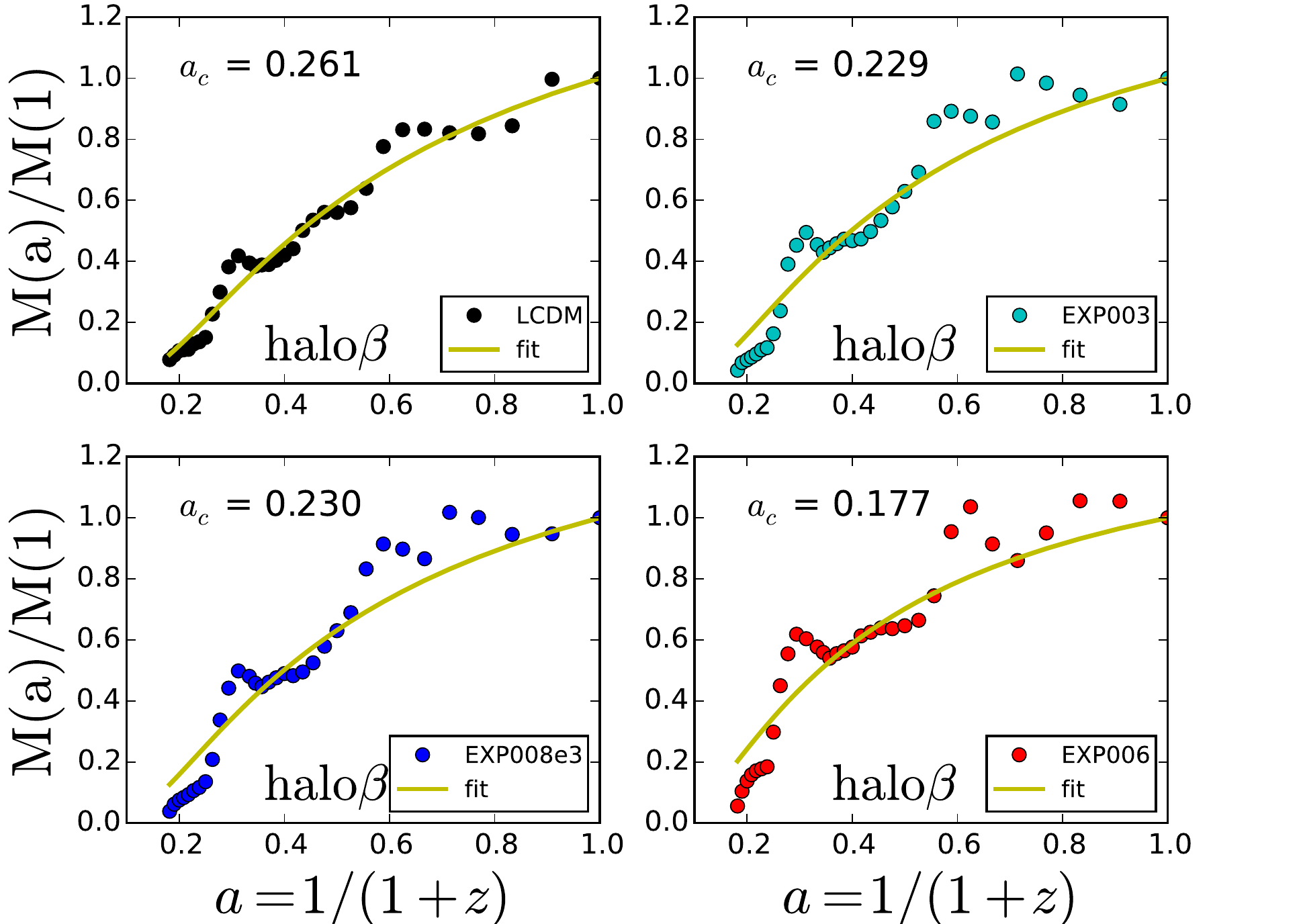,width=0.52\textwidth}
\caption{Accretion histories for halo$\beta$, where we show for each panel a different cosmological model. The fit to the accretion histories of each realization (solid yellow line) is obtained from Eq.~\ref{eq:acc}.}
\label{fig:accfit}
\end{figure}

In order to estimate the time of formation for each halo, we followed the approach described in \citet{Wec02}. In their paper they show how halo accretion histories are well described by an exponential form that depends on one parameter, the formation epoch $a_c$, which is defined as the expansion factor at which the logarithmic derivative of the mass evolution falls below a critical value $S$. Specifically, the fitting form is given by
\begin{align}
\label{eq:acc}
\frac{M(a)}{M(1)} = \exp \left[-a_c S \left(\frac{1}{a}-1\right)\right]
\end{align}
with $S=2$. We used the fitting function of Eq.~\ref{eq:acc} to compute the formation epochs for each of the three halos for all cosmologies. As an example, we show in Fig.~\ref{fig:accfit} the fitting result for halo$\beta$ and we summarize the formation epochs for all halos in Table~\ref{table:acc}. 
\begin{table}
\centering
\caption{Values for the formation epochs $a_c$ for halo$\alpha$, halo$\beta$, halo$\gamma$ in all four cosmologies.}
\begin{tabular}{@{}cccccccr@{}}
  \hline 
 				& \LCDM		& EXP003 	& EXP008e3 	& EXP006\\ 
  halo$\alpha$ 		& 0.194  		& 0.181		& 0.183		& 0.142    \\ 
  halo$\beta$ 		& 0.261  		& 0.229		& 0.230		& 0.177    	\\ 
  halo$\gamma$ 	& 0.225  		& 0.210		& 0.215		& 0.173    \\ 
  \hline   
\end{tabular}
\label{table:acc}
\end{table}

As pointed out in \citet{Wec02}, \citet{Dut14} and \citet{Lud13}, in a \LCDM cosmology an early formation epoch leads to higher concentrations. The same happens for dynamical dark energy cosmologies, e.g. \citet{Kly03,Dol04}. Interestingly, in coupled dark energy cosmology this behavior is not preserved. Despite the fact that a stronger coupling brings to an earlier halo formation epoch, halo concentrations decrease when the coupling is increased, as clearly visible from the EXP006 realizations. This phenomenon has also been showed in \citet{Bal10} and \citet{Bal11b}, where an analysis of the modifications due to the coupling has been extensively carried out. The result is that the friction term in Eq.~\ref{linper} is responsible for making the halo expand by altering its virial equilibrium through the injection of kinetic energy in the system, which in turns lowers the concentration \citep[see][]{Bal11b}. This shows how in coupled cosmologies the lower concentrations are not the result of formation histories but the effect of the modified dynamics. 
\subsection{Subhalos}
\label{sat}
\quad\quad In this Section we study the subhalo abundance, their radial distribution and circular velocities.
\subsubsection{Abundance}
\begin{figure}   
\psfig{file=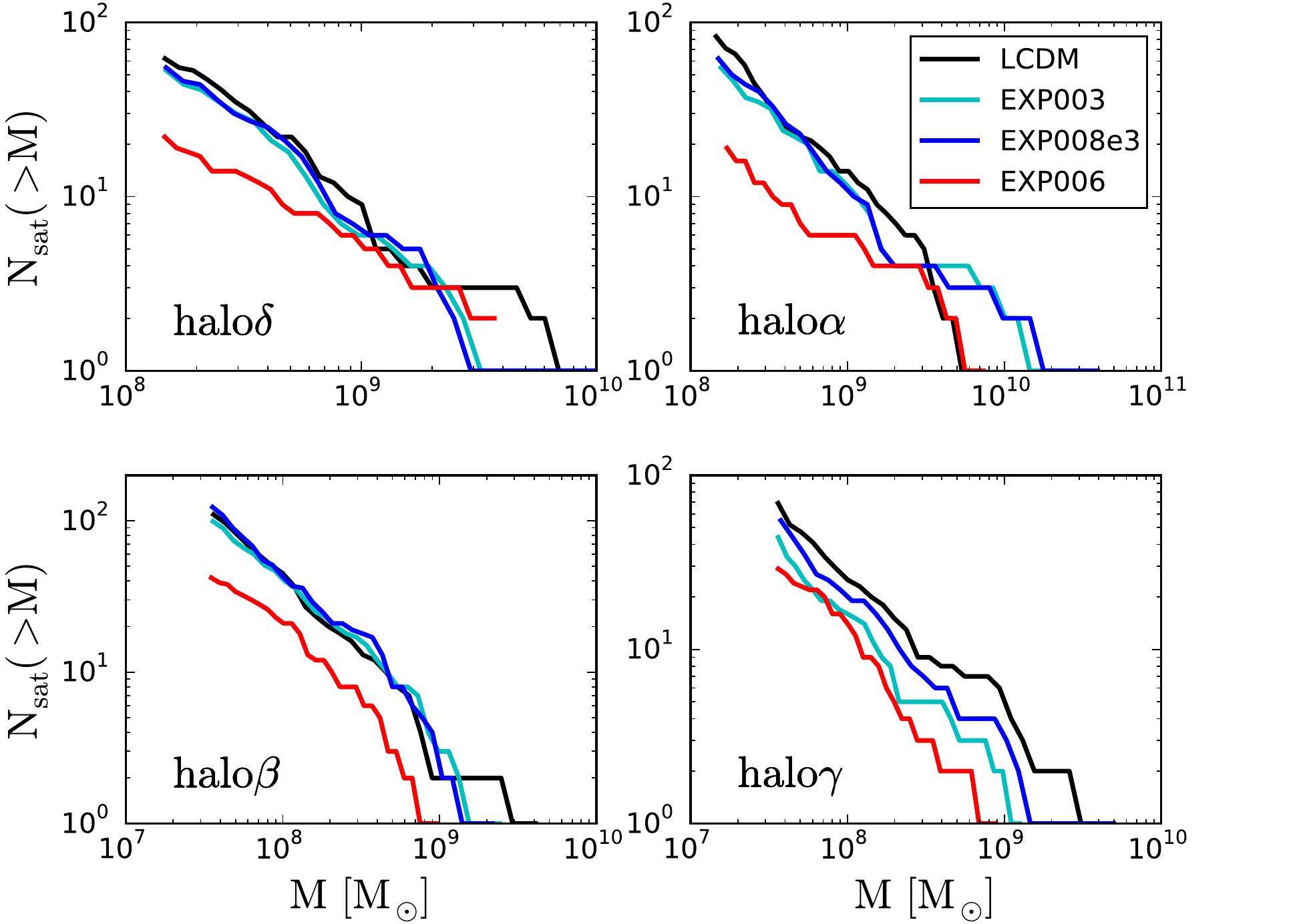,width=0.52\textwidth}
\caption{Cumulative number of subhalos with more than 400 particles as function of their mass for halo$\alpha$, halo$\beta$, halo$\gamma$ and halo$\delta$ at $z=0$ for each cosmology.}
\label{fig:abund}
\end{figure}

\quad\quad The lower number of substructures present in EXP006 halos compared to \LCDM can be recognized in Fig.~\ref{fig:maps}. In Fig.~\ref{fig:abund} we show the subhalo mass function, where only subhalos that lie within R$_{200}$ and that have more than 400 particles are considered. The total number of subhalos in EXP006 realizations is always from 50 to 75\% lower than in the respective \LCDM cases, while differences between EXP003 and EXP008e3 and \LCDM are much less evident ($\sim$ 10\%). Thus, the missing satellites problem \citep{Kly99,Moo99} can be progressively alleviated when increasing the coupling. Note that the differences in the subhalos minimum mass among halo$\alpha$, halo$\beta$ and halo$\gamma$, halo$\delta$ are due to the different resolutions used (see Section~\ref{nbodysim}).
\subsubsection{Radial Distribution}
\begin{figure}   
\psfig{file=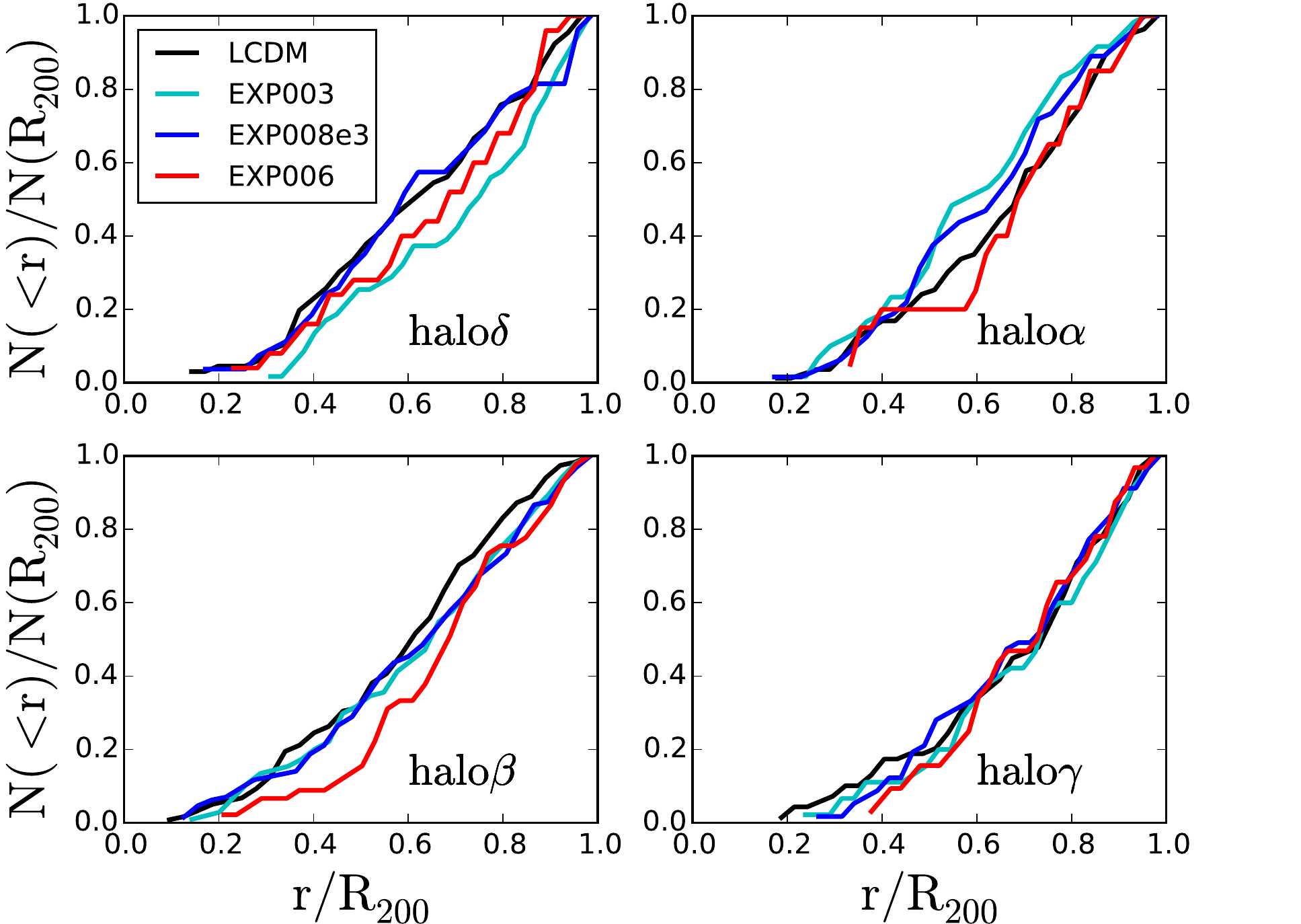,width=0.52\textwidth}
\caption{Cumulative number of subhalos with more than 400 particles as function of distance from the main halo center for halo$\alpha$, halo$\beta$, halo$\gamma$ and halo$\delta$ at $z=0$ for each cosmology.}
\label{fig:num_dist_cumul}
\end{figure}

\begin{figure}   
\psfig{file=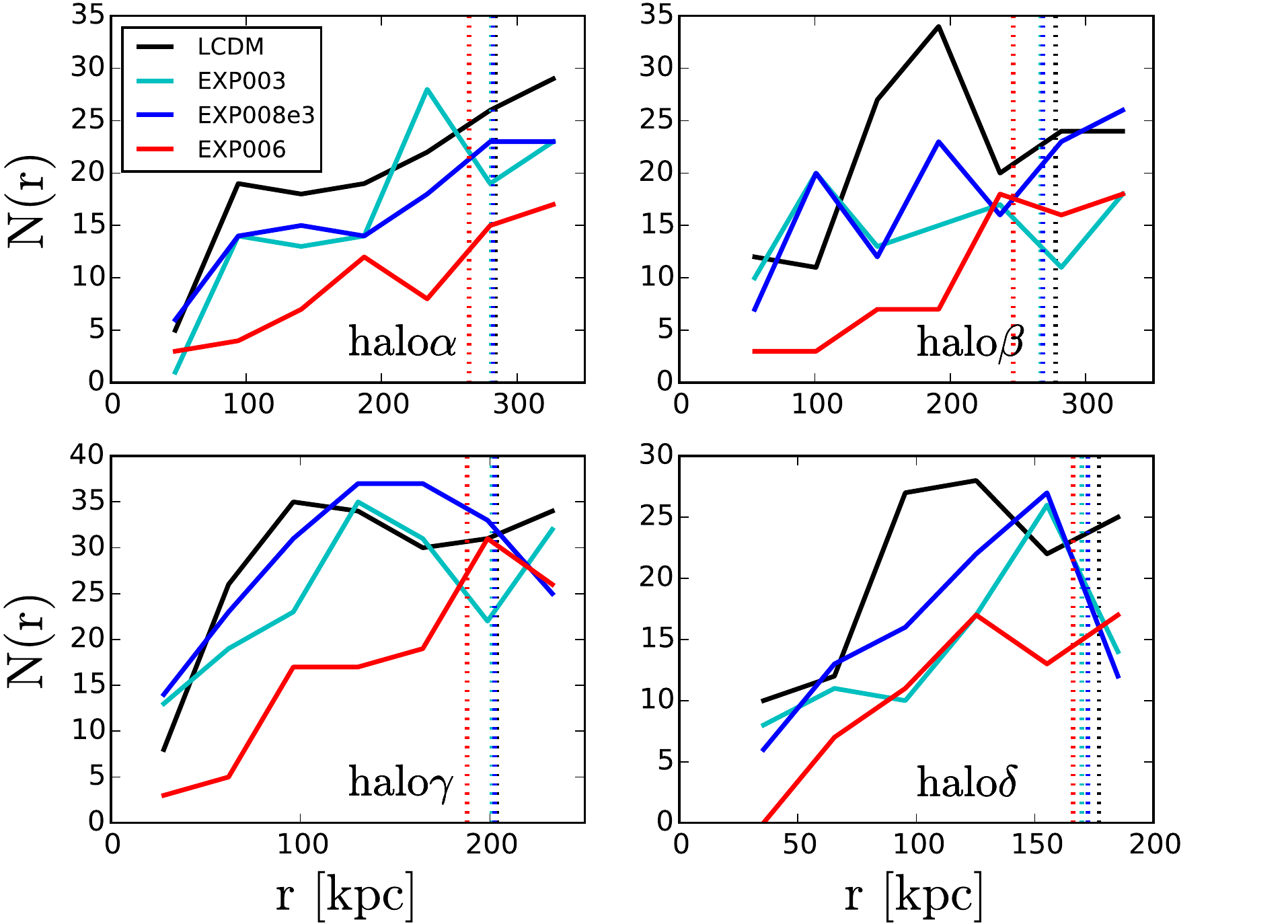,width=0.52\textwidth}
\caption{Differential number of subhalos in R$_{200}$ with more than 400 particles as function of distance from main halo center for halo$\alpha$, halo$\beta$, halo$\gamma$ and halo$\delta$ at $z=0$ for each cosmology. The vertical dashed lines mark R$_{200}$ for each halo in each cosmology. For a given halo, the binning is kept constant for all cosmologies.}
\label{fig:num1rvir}
\end{figure}

\quad\quad Fig.~\ref{fig:num_dist_cumul} shows the cumulative distribution of subhalos as a function of the distance from the main halo center normalized to the total number of subhalos within R$_{200}$. 
All halos in all cosmologies show non-significant differences in the cumulative radial distribution. In order to better understand the distribution of subhalos, in Fig.~\ref{fig:num1rvir} we show the differential distribution in a sphere of constant radius for all cosmologies, the radii are 350 kpc for halo$\alpha$ and halo$\beta$, 250 kpc for halo$\gamma$ and 200 kpc halo$\delta$. 
The number of bins is kept the same for each halo in all cosmologies and the vertical lines show the virial radii. The distributions show a clear decrease of the number of subhalos in EXP006 halos compared to their respective \LCDM cases, while for EXP003 and EXP008e3 cosmologies differences are not so evident. 

\begin{figure}   
\psfig{file=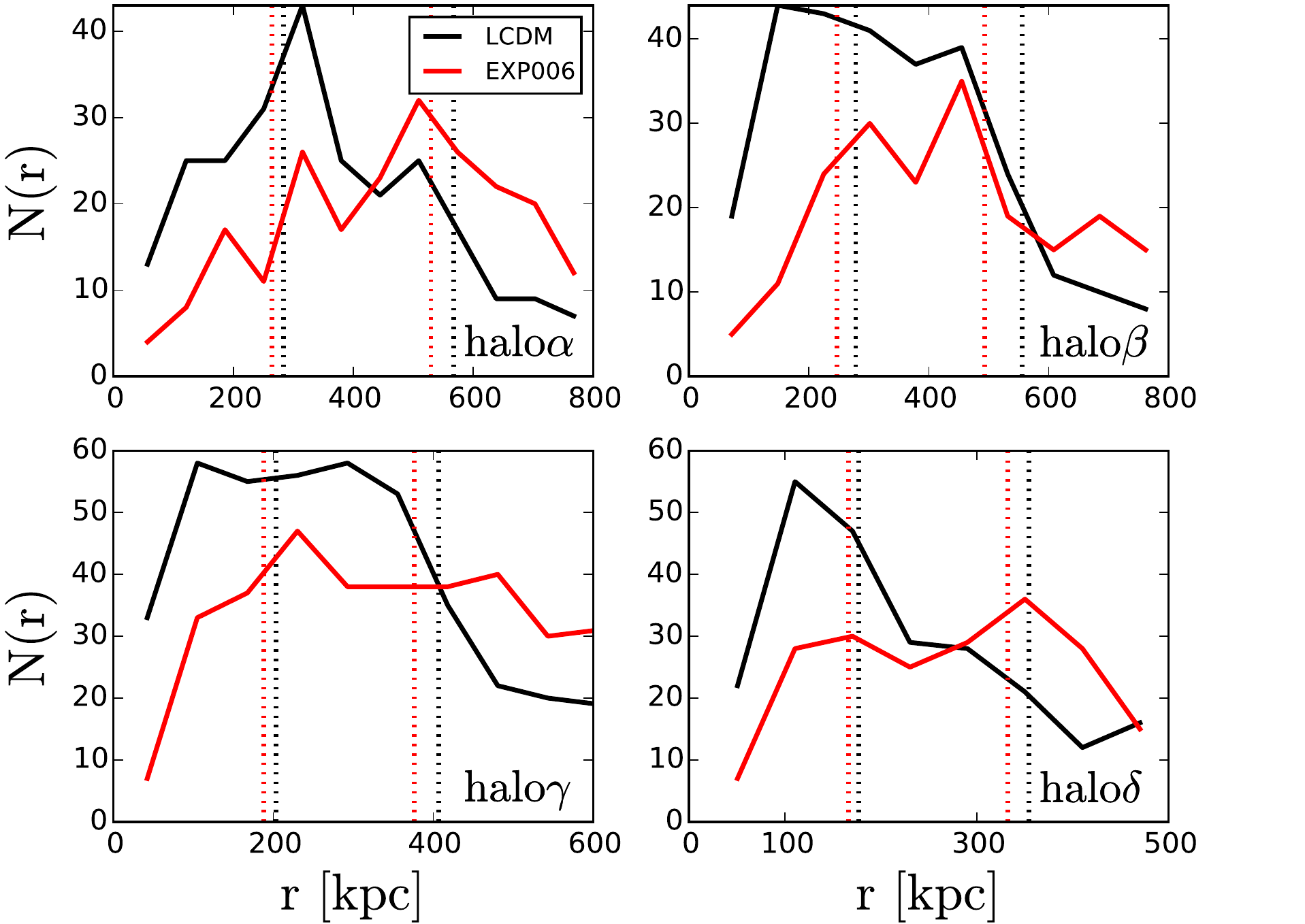,width=0.52\textwidth}
\caption{Number of subhalos in 3R$_{200}$ with more than 400 particles as function of distance from main halo center for halo$\alpha$, halo$\beta$, halo$\gamma$ and halo$\delta$ at $z=0$ for each cosmology. The vertical dashed lines mark R$_{200}$ and 2R$_{200}$ for each halo in each cosmology. For a given halo, the binning is kept constant for all cosmologies.}
\label{fig:num3rvir}
\end{figure}

As pointed out in Section~\ref{acc_hist}, coupled cosmologies decrease halo concentrations thanks to the presence of a friction term in the equation for the evolution of density perturbations (Eq.~\ref{linper}), despite the earlier halo formation epochs. We claim that the same effect can be responsible for the lower number of subhalos compared to $\Lambda{\rm CDM}$. Thanks to the extra friction term and to subhalo lower concentrations, subhalos that are falling into the main halo potential well are more heavily stripped, thus less halos with more than 400 particles survive. If this claim is correct, we should be able to find a difference in the subhalos number distribution when we reach distances from the main halo center that are bigger than the radius from which the gravitational influence of the host halo is felt. In Fig.~\ref{fig:num3rvir} we show the differential radial distribution of the number of subhalos out to about three times the virial radius of each halo. 
For the sake of clarity, we choose to show only the most extreme cases, EXP006, and \LCDM for all four halos. The dotted lines represent one and two times R$_{200}$ for each halo in each cosmology. What we would like to stress, is that there seem to be a decrease in the number of subhalos in halos living in EXP006 cosmology compared to their \LCDM realizations \emph{only} within the gravitational influence of the main halo. Between 1.5 and 2 R$_{200}$ this behavior inverts and halos living in the strongly coupled cosmology seem to have a larger or at least a comparable number of subhalos with respect to their \LCDM cases. Thus, we ascribe the presence of a lower subhalos number to a massive stripping effect rather than EXP006 producing intrinsically a lower number of subhalos. We would like to stress on the fact that lowering the number of subhalos can also be achieved by warm dark matter cosmologies (e.g. \citealt{And13}), but the fundamental difference lies on the fact that those subhalos in warm dark matter cosmologies were never formed, while in coupled dark energy cosmologies subhalos do form but they are heavily stripped.
\subsubsection{Circular Velocities}
\quad\quad \citet{Boy11} first showed that N-body simulations of a Milky-Way size halo predict a significant number of subhalos with circular velocities higher than the circular velocities that we measure for the brightest satellites of the Milky Way, which is surprising since these massive subhalos should not fail in producing stars.

The discrepancy between \LCDM prediction and observations can be alleviated in multiple ways, starting from baryonic processes. \citet{Bro14} suggest that baryonic feedback processes could be responsible for a dark matter redistribution, with the result of decreasing the central densities of the most massive subhalos. \citet{Ras12} point out that the possibility of star formation being stochastic below a certain mass would justify the Milky Way having massive dark satellites; furthermore, they highlight the fact that the tension between the Via Lactea II simulation and observations is only a factor of two in mass, which suggests that the uncertainty on the Milky Way virial mass could be a viable way out from the tension \citep{Ver13,Ken14}. \citet{Pur12} showed that there exists a significant variation in subhalo properties even when the host halos have the same virial mass. 

Last but not least, the discrepancy can be alleviated by appealing to non-\LCDM cosmologies. The cases for warm, mixed (cold and warm) and self-interacting dark matter are considered respectively in \citet{Lov12}, \citet{And12,And13}, \citet{Vog12}. In all cases they find that subhalos are less concentrated due to their late formation time, suggesting that alternative cosmologies can contribute to alleviate the tension between predictions and observations.

\begin{figure*}   
\centering
\begin{minipage}{180mm}
\psfig{file=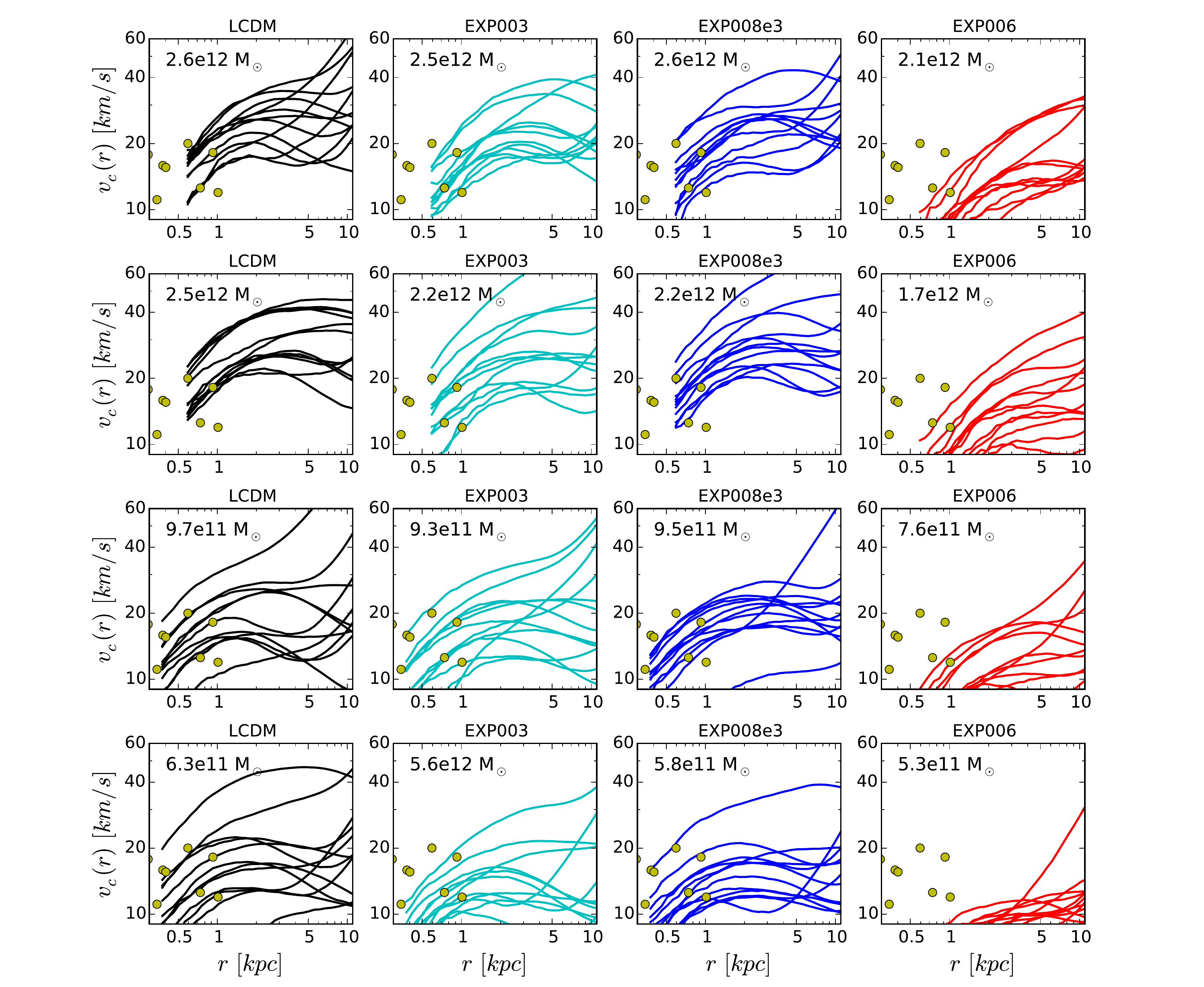,width=1.0\textwidth}
\caption{Rotation curves of the most massive subhalos at the moment of infall for each halo in each cosmology. From the top row down we show halo$\alpha$, halo$\beta$, halo$\gamma$, halo$\delta$, from left to right we show $\Lambda$CDM (black), EXP003 (cyan), EXP008e3 (blues), EXP006 (red). We estimate the subhalo mass ranking at the moment of infall using the correlation between orbital energy and subhalo mass loss found in \citet{And13}. The yellow points are the observed values for $v_{\rm circ}(r_{1/2})$ for the brightest dwarf galaxies orbiting around the Milky Way. Data are taken from \citet{And13} and references therein. The masses of each main halo realizations is written on each panel.}
\label{fig:curves_sat}
\end{minipage}
\end{figure*}

In Fig.~\ref{fig:curves_sat} we show the rotation curves for the twelve most massive subhalos at the moment of infall. We used the correlation between orbital energy and subhalo mass loss found in \citet{And13} to determine the subhalos ranking. 
Each row illustrates the twelve subhalos rotation curves for a given main halo in all considered cosmologies, from the top down we show halo$\alpha$, halo$\beta$, halo$\gamma$, halo$\delta$. In yellow we show the observed values for $v_{\rm circ}(r_{1/2})$ for the brightest dwarf galaxies orbiting around the Milky Way, data are taken from \citet{And13} and references therein. Despite halo$\alpha$ and halo$\beta$ having comparable masses, the tension between simulated curves and measured points in the \LCDM case is more evident in halo$\beta$, supporting the fact that subhalo properties can vary even when host halos have the same virial mass \citep{Pur12}. The tension is alleviated in the case of halo$\gamma$ and even more halo$\delta$, given their lower masses \citep{Ver13}. Overall, when looking at all halos in EXP003 and EXP008e3 cosmologies these do not show significant improvement compared to their \LCDM realizations in decreasing the inner densities of subhalos. On the other hand, in the case of EXP006 cosmology, all four halos show such a dramatic decrease in subhalos rotational velocity peaks that rotation curves become incompatible with measured values. The dramatic decrease was to be expected given the choice of a large coupling parameter for EXP006 cosmology, but nonetheless it is useful to understand the effects of the coupling.
\subsection{Zooming-in on a dwarf halo}
\label{dwarfsection}
\begin{figure}   
\centering
\psfig{file=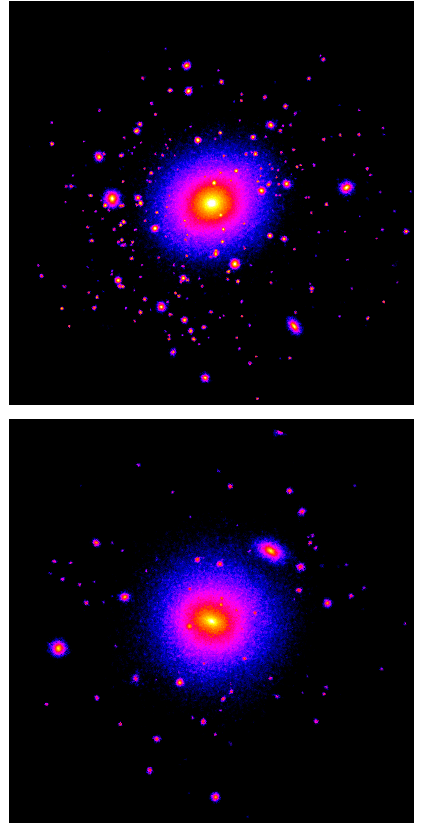,width=0.45\textwidth}
\caption{Density maps for halo$\epsilon$ in \LCDM (upper panel) and in EXP006 cosmology (lower panel). The side of each projection is 2$\times$R$_{\rm 200}$.}
\label{dw:maps}
\centering
\end{figure}

\begin{figure}   
\centering
\psfig{file=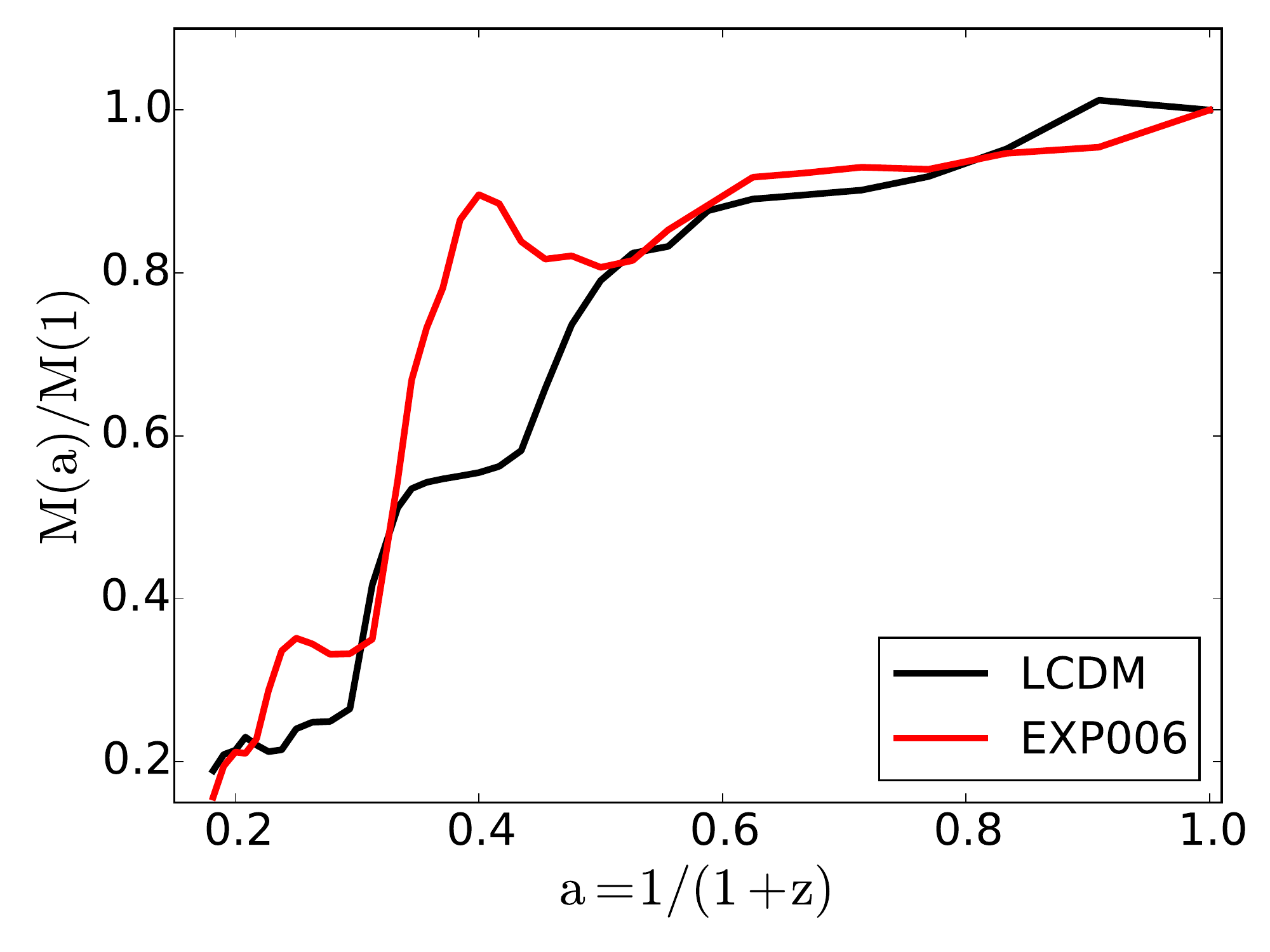,width=0.35\textwidth}
\caption{Accretion histories for halo$\epsilon$ in \LCDM (black) and in EXP006 (red) cosmology. We show the ratio of M$_{\rm 200}$ at a given time and M$_{\rm 200}$ today as a function of the scale factor.}
\label{dw:acc}
\end{figure}

\quad\quad To better explore the effects of the coupling at high resolutions, we simulated a dwarf galaxy halo, halo$\epsilon$. We chose an isolated halo (no structures with comparable mass within four of its virial radii) and, given the results of Section~\ref{sat}, we only focused on the two most distant cosmological cases, \LCDM and EXP006 cosmology. The virial masses are respectively 4$\times$10$^9$M$_{\odot}$ and 3$\times$10$^9$M$_{\odot}$, with a mass resolution of 1.3$\times 10^{3}$M$_{\odot}$. We show in Fig.~\ref{dw:maps} the density maps for halo$\epsilon$ in both cosmologies, \LCDM on the upper panel. It is visible how the number of substructures decreases in the case with coupling. 
\begin{figure}
\psfig{file=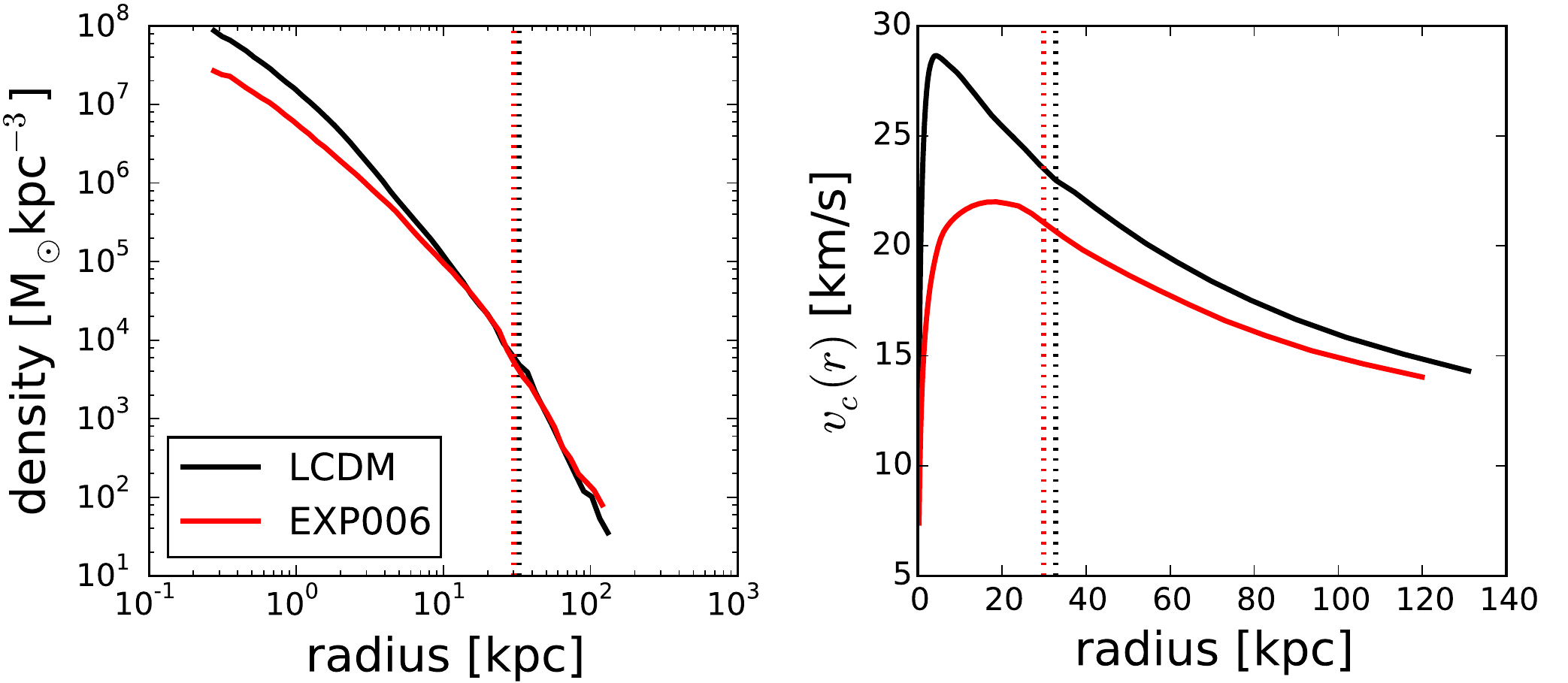,width=0.5\textwidth}
\caption{Density profiles and rotation curves for halo$\epsilon$ in \LCDM (black) and in EXP006 (red) cosmology. The inner minimum radius is three times the softening length, while the most outer radius is four times R$_{\rm 200}$ respectively. The vertical dashed lines represent R$_{\rm 200}$ for each cosmology.}
\label{dw:profvel}
\end{figure}
In Fig.~\ref{dw:profvel} we show density profiles and rotation curves for halo$\epsilon$ in both cosmologies. The effect of the coupling is very evident in lowering the concentration and flattening the rotation curve. The values for the halo concentrations are $c=15.2$ and $c=6.5$ for \LCDM and EXP006 cosmology respectively. Although the density profile in the coupled dark energy case is less concentrated, it is still cuspy, showing that in coupled cosmologies, as in $\Lambda$CDM, we are not able to produce a dark matter only cored density profile. The inconsistency with observation thus still persists, given the observational evidence that supports cored density profiles for the satellites of the Milky Way \citep{Wal11,Amo12,Amo13}. Interesting to note, by constructing a model in which both warm and cold dark matter are present and only the cold component is coupled to dark energy, a very high value ($\beta_c\sim 10$) for the coupling constant is favored \citep{Bon15} and simulated dark matter only dwarf halos show a cored density profile \citep{Mac15}. Fig.~\ref{dw:acc} shows accretion histories, the ratio of M$_{\rm 200}$ to its value today is plotted as function of scale factor. As in Section~\ref{acc_hist}, we calculated the formation epochs as in \citet{Wec02} and obtained $a_{c} = 0.191$ ($\Lambda$CDM), $a_{c} = 0.146$ (EXP006), confirming the finding that coupled dark energy models have earlier formation times, also for less massive halos. 
\section{Conclusions}
\label{conclusions}
\quad\quad We have performed the first study in coupled dark energy cosmologies on high resolution simulations on galactic scales, with the aim to study the effects of the coupling between dark energy and dark matter on these scales, so far neglected in previous studies. We chose to investigate two viable models, one with constant coupling and one with varying coupling with redshift; we also chose a third case where the constant coupling value has been pushed beyond observational constraints to better investigate its effects. We then selected three Milky-Way size halos, a 6$\times$10$^{11}$M$_{\odot}$ halo and a dwarf halo 5$\times$10$^{9}$M$_{\odot}$, and studied their properties in a \LCDM reference model and in the coupled cosmologies, resolving each halo with $\sim 10^6$ particles.

We computed concentrations and formation epochs for all halos and we find that, despite the earlier formation epochs of the coupled cosmologies halos, these have lower concentrations. In a \LCDM or a dynamical dark energy scenario, earlier formation epochs would imply higher concentrations, but in the coupled dark energy case the reason for lower concentrations is not related to formation histories, but rather to the modified dynamics. In fact, the equation for the linear evolution of density perturbations (Eq.~\ref{linper}) shows the friction term $-\beta\dot\phi\dot\delta_c$, an extra term compared to the \LCDM case that redistributes the dark matter particles and lowers the central densities, despite the earlier formation times (see \citealt{Bal11b}). We find that this behavior is reproduced for all mass scales that we have investigated.

In particular, subhalos can also be significantly less concentrated. When falling towards their host, they are more heavily stripped once they start feeling the gravitational influence of the host halo. This translates into decreasing the number of subhalos compared to the \LCDM realization and, additionally, subhalos are themselves less massive and less concentrated. For these reasons, coupled cosmologies can be helpful in alleviating satellite-scales inconsistencies of $\Lambda$CDM. On the other hand, we find that in order to try to solve these issues with the coupling alone, one needs to use an extreme value for the coupling constant that is ruled out by observational constraints. In fact, only in the case with the highest coupling value the number of subhalos is significantly reduced (up to 75\% less subhalos) than in the respective \LCDM cases, while for the viable coupling cosmologies the decrease is much less significant (10\% less subhalos). Moreover, we find that the distribution of the subhalos inside the main halo virial radius does not vary significantly among cosmologies. Lastly, less concentrated coupled cosmologies subhalos can in principle be useful to reconcile the inconsistency between the observed properties of the Milky Way dwarf galaxies and \LCDM simulations predictions, but once more a high enough value for the coupling must be assumed. Interestingly, allowing the introduction of massive neutrinos does alleviate the constraints on the coupling \citep[see e.g.][]{LaV09}, leaving coupled dark energy models dynamics on sub-galactic scales an interesting option.

Overall coupled dark energy models can be as effective as \LCDM in reproducing observations on sub-galactic scales and, for specific choices of the coupling, they can improve the agreement between predicted and observed properties. Hence, coupled models would need to be further investigated, possibly taking into account the effects of baryons at sub-galactic scales, which, as already shown in dynamical dark energy models \citep{Pen14}, are expected to amplify differences observed in the dark matter only case.


\section*{Acknowledgments} 

The numerical simulations were performed on the THEO cluster of the Max-Planck-Institut f\"ur Astronomie and Hydra cluster, both based at the Rechenzentrum in Garching. CP and AVM acknowledge the support from the Sonderforschungs-bereich SFB 881 ÒThe Milky Way SystemÓ (subproject A02) of the German Research Foundation (DFG). CP also acknowledges the support of the International Max Planck Research School, Heidelberg (IMPRS-HD). MB acknowledges partial support by the  Marie Curie Intra European Fellowship
``SIDUN"  within the 7th Framework Programme of the European Commission. LC acknowledges the Brazilian research Institutions FAPES and CNPq for financial support.

\bibliographystyle{apj} \bibliography{biblio}

\bsp
\label{lastpage}
\end{document}